\documentclass[preprint]{aastex}
\begin{document}

\title{Searching for Planets in the Hyades III: The Quest for Short-Period 
Planets
\footnote{Some data were obtained with the HET. The Hobby-Eberly 
Telescope is operated by McDonald Observatory on behalf of The University of Texas at Austin, the
Pennsylvania State University, Stanford University,
Ludwig-Maximilians-Universit\"{a}t M\"{u}nchen, and
Georg-August-Universit\"{a}t G\"{o}ttingen.}\ $^{,}$
\footnote{Additional data presented herein were obtained at the 
W.M. Keck Observatory, which is operated as a scientific partnership among
the California Institute of Technology, the University of California and 
the National Aeronautics and Space Administration. The Observatory was made
possible by the generous financial support of the W.M. Keck Foundation.}}

\author{Diane B. Paulson\altaffilmark{3}}
\affil{Department of Astronomy, University of Texas, Austin, TX 78712}
\email{apodis@astro.as.utexas.edu}

\altaffiltext{3}{current address: Department of Astronomy, University of
Michigan, 830 Dennison, Ann Arbor, MI 48109}

\author{Steven H. Saar}
\affil{Center for Astrophysics, 60 Garden Street, Cambridge, MA 02138}
\email{ssaar@cfa.harvard.edu}

\author{William D. Cochran}
\affil{McDonald Observatory, University of Texas, Austin, TX 78712}
\email{wdc@astro.as.utexas.edu}

\and

\author{Gregory W. Henry\altaffilmark{4}}
\affil{Center of Excellence in Information Systems, Tennessee State University, 
Nashville, TN 37203}
\email{henry@schwab.tsuniv.edu}

\altaffiltext{4}{Also Senior Research Associate, Department of Physics and 
Astronomy, Vanderbilt University, Nashville, TN  37235}

\begin{abstract}
We have been using the Keck I High Resolution Spectrograph (HIRES)
to search for planetary companions in the Hyades cluster. We selected four stars 
from this sample which showed significant radial velocity variability 
on short timescales to search for short-period planetary companions. The radial velocities of these four stars were monitored regularly
with the Hobby Eberly Telescope (HET) for approximately two months, while 
sparse data were also taken over $\sim$4 months:
we also obtained near-simultaneous photometric observations
with one of the automatic photoelectric telescopes at Fairborn Observatory. 
For three of the stars, we detect photometric 
variability with the same period present in the radial velocity ($v_{\rm r}$) 
measurements, compatible with the expected rotation rates for Hyades members. 
The fourth star continues to show $v_{\rm r}$ variations and minimal 
photometric variability but with no significant periodicity. This study shows 
that 
for the three stars with periodic behavior, a significant portion of 
the $v_{\rm r}$ fluctuations are  
likely due primarily to magnetic activity modulated by stellar rotation rather than planetary companions. 
Using simple models for the $v_{\rm r}$ perturbations arising from 
spot and plage, we demonstrate that {\it both} are likely to contribute to 
the observed $v_{\rm r}$ variations.  
Thus, simultaneous monitoring of photometric (photospheric) and 
spectroscopic (chromospheric) variations is essential 
for identifying the cause of Doppler shifted absorption lines in more active 
stars.
\end{abstract}


\section{Introduction}

It is well known that starspots will cause shifts in line profile shapes 
\citep[cf.][]{VoPeHa87}, which will cause apparent
Doppler shifts of the lines. \citet{SaDo97} modeled the expected $v_{\rm r}$ 
variability due to the rotational modulation of spots. 
\citet{Ha99} and \citet{Ha02} make similar calculations and obtain the 
same results as \citet{SaDo97}. Recently, observational
data to support this has been published. \citet{QuHeSi01} found
HD~166435 to have a large velocity amplitude, 
but it turned out also to show 
sinusoidal photometric amplitude in Str\"omgren $y$ with the
same period. In 
addition, the Calcium HK index varied smoothly on the same timescale. This
indicated the 
$v_{\rm r}$ variations were most likely due to stellar activity rather than a planetary companion. 
\citet{HeDoBa02} showed starspots to be the cause of the line shifts in 
HD 192263
and GWH and P. Butler see the same thing in HD 19632 (private communication, 
2002). Additionally, \citet{SaBuMa98} and \citet{SaMaNa00a} confirm the 
models presented by \citet{SaDo97}.

The amplitude of $v_{\rm r}$ caused by plage regions is also 
beginning to be modeled \citep{Sa03}. He finds that this can be several 
tens of m s$^{-1}$. Therefore, it is important that one monitor all aspects
of stellar activity when searching for planets, particularly
around active stars.    
Here, we investigate the implications of the rotational modulation of stellar 
activity on our search for short-period planets in the Hyades.

\section{Observations and Analysis}

\subsection{Sample}

The motivations for the Keck Hyades survey are discussed by \citet{CoHaPa02}.
With a sample of 98 stars, with [Fe/H]=0.13 \citep{PaSnCo03}, we might 
expect a small number of short-period planets similar to 51 Peg \citep{MaQu95}.
From our Keck Hyades sample, we chose four 
stars for follow-up observations with the HET's High Resolution Spectrograph 
(HRS) in search of 
short-period planetary companions. The targets were chosen from a group of stars
which showed 
significant $v_{\rm r}$ rms on short timescales. Only
four were observed at this time due to HET scheduling constraints.
The image quality of the telescope was being corrected during 
this time, and though this did not affect the quality of the observations, 
it placed a magnitude limit on the stars which could be observed. The observed targets are 
listed in Table 1. 
\subsection{$v_{\rm r}$ measurements}

The $v_{\rm r}$ observations from the Keck High Resolution Echelle 
Spectrograph (HIRES) are described in full in \citet{CoHaPa02}. 
We began regular observations of these four stars from late-December 2001 to
Mid-February 2002 (and a few observations of each star taken sporadically during 
the fall of 2001) with the HRS at the HET \citep{Tu98, CoTuMc03}. 
We used the 3" optical fiber 
feed to the HRS with resolving power R=60,000. We set the 316 g/mm cross 
disperser to central wavelength 5938\AA. This includes almost the entire 
I$_{2}$ region ($\approx 5000 - 6200$\AA; I$_{2}$ is used as the velocity metric), on one of the CCD chips. Any other 
configuration would result in spreading the I$_{2}$ region over both
CCD chips and losing some I$_{2}$ information in the gap between CCDs. We 
manufactured the I$_{2}$ gas absorption cell for use in the HRS at the 
University of Texas, and during these observations, it was run at 60$^{\circ}$C.

Each exposure was restricted to 15 minutes in length 
to reduce velocity smearing due to the Earth's rotation. 
This limited the 
signal-to-noise ratio (S/N), but was necessary to obtain high $v_{\rm r}$ precision (see \nocite{PaSaCo02}Paulson et al. 2002).
The S/N varied greatly from exposure to exposure due to 
seeing variations, but all observations had S/N $\ga$200 per pixel.
The CCD images were reduced and extracted using standard IRAF\footnote{IRAF is distributed by the National Optical Astronomy Observatories,
    which are operated by the Association of Universities for Research
    in Astronomy, Inc., under cooperative agreement with the National
    Science Foundation.} packages.
We use a program called RADIAL (developed at the University of 
Texas (UT) and McDonald Observatory) to measure precise 
radial velocities. This program was adapted for use with data 
from all of the planet search programs affiliated with UT. Brief discussions 
of the program may be found in \citet{CoHaBu97} and \citet{HaCoMA00}. 
The typical velocity precision for observations with the HET HRS is 
4-6~m~s$^{-1}$ \citep{CoTuMc03}.
The $K$ amplitudes are listed in Table 1.
The $K$ amplitude for HD~26756, which is not found to have significant 
periodicity,
is just defined as one half of the peak-to-peak variation of the $v_{\rm r}$.
The $v_{\rm r}$ data are listed in Table 2.

\subsection{Photometric measurements}

Between 11 and 14 nightly photometric observations of each of the four stars 
were acquired in 2002 February and March with the T12 0.8~m automatic 
photoelectric telescope (APT) at Fairborn Observatory in the Patagonia 
Mountains of southern Arizona\footnote{Further information about Fairborn 
Observatory can be found at \url{http://www.fairobs.org/}.}.  The T12 APT 
measures the difference in brightness between a program star and nearby 
comparison stars in the Str\"{o}mgren $b$ and $y$ passbands.  The observing 
procedures and data reduction techniques employed with this APT are identical 
to those for the T8 0.8~m APT described in \citet{He99}.  The resulting 
Str\"{o}mgren $b$ and $y$ differential magnitudes were corrected for 
differential extinction with nightly extinction coefficients and transformed 
to the Str\"{o}mgren system with yearly mean transformation coefficients.  The 
external precision of the differential magnitudes, defined as the standard 
deviation of a single differential magnitude from the seasonal mean of the 
differential magnitudes, is typically around 0.0012 mag for this telescope, 
as determined from observations of pairs of constant stars.  Our primary 
comparison stars for were HD~26737 (for HD~26736), HD~27561 (for HD~26756 and 
HD~26767), and HD~18579 (for HIP~13806); all three comparison stars are 
constant 
to $\sim$0.003 mag or better as determined by intercomparison with additional 
comparison stars.  The resulting range in the differential $y$ magnitudes 
($\Delta y$) of our four program stars are given in Table~1.  Photometric data 
are listed in Table 3.

\section{Results}

\subsection{$P_{\rm rot}$}
We are able to determine the rotational period ($P_{\rm rot}$)
for these stars from both sets of observations- $v_{\rm r}$ and 
$\Delta y$, independently. We 
used the method of \citet{HoBa86} for period determination, and all results 
are listed in Table 1. 
We independently determined periods from the photometric data using the 
procedure 
outlined in \citet{HeFeKa01}, and these periods are as follows: HIP~13806- 
9.57$\pm$0.18 d, HD~26736- 8.48$\pm$0.35 d, HD~26767- 8.69$\pm$0.13 d. These 
agree with the periods determined by the method of Horne \& Baliunas (listed
in Table 1).
HD~26736, HD~26767 and HIP~13806
all show relatively significant  periods in the period analysis, with false 
alarm probabilities (FAPs) of 
$\lesssim 15\%$ (also calculated by the method described in Horne \& Baliunas).
$P_{\rm rot}$ is also consistent 
between spectroscopic and photometric data assuring us that the periods derived 
in $v_{\rm r}$ are due to rotational modulation of stellar activity. We  
chose to use $P_{\rm rot}$ determined from $v_{\rm r}$ to show the phase 
plots in Figure 1 since there were more data available. We could, equally as 
well, have chosen to phase the plots according to $P_{\rm rot}$ as 
derived from the photometry.  Zero phase was chosen to be at 
photometric maximum, which roughly corresponds to the time at which the 
$v_{\rm r}$ curve crosses from ``blue'' to ``red''. This is done so because 
photometric maximum is an easily understandable physical event, thus
``grounding'' the phase plot. The associated 
$v_{\rm r}$ plots are only the results of the active regions (photometric 
variations). 

We searched the Hipparcos Epoch Photometry \nocite{ESA}(ESA) of each of these 
stars for periodicity. The period finding algorithm of Horne 
\& Baliunas 
was again used to search for periodicity in the Hipparcos data 
set. When \citet{QuHeSi01} studied HD~166435, they noticed that by taking 
small sections 
of their data in time, they were able to recover $P_{\rm rot}$. However, 
when taking the entire data se set, they saw no obvious signature of 
$P_{\rm rot}$. This was an effect of phase shifts as the spots migrated 
either in longitude with stellar differential rotation or in
latitude $or$ appeared or disappeared asymmetrically. HD~26736 and HD~26767
appear to behave like HD~166435 in that the full photometric data set from 
Hipparcos shows no obvious periodicity.
This is also true for the 
$v_{\rm r}$ data sets of these stars observed from Keck, although weak signals 
may be present for these two stars. However, the $P_{\rm rot}$ is recovered when taking small
intervals of Keck $v_{\rm r}$ data in time for these two program stars. 
The case of HIP~13806 is different; we found a strong period at 
9.60 days (FAP of 0.02\%) in the full Hipparcos data set. The 
periodogram peak 
is sufficiently wide that it encompasses $P_{\rm rot}$ derived from both 
$v_{\rm r}$ and $\Delta y$. Thus, $P_{\rm rot}$ is somewhat poorly defined from the 
Hipparcos observations alone.
The phase curve for the HIP~13806 Hipparcos data with a 9.42 day period is 
shown in Figure 2. HIP~13806 is unusual in that the same period 
seems to be more or less coherent, though quite noisy, over the $\sim$2.5 year 
time frame of 
the Hipparcos observations. HIP~13806 is also notable 
in that it is one of few dwarf stars which show this long term stability of 
active regions. \citet{ToGr88} also observed the G8 dwarf star $\xi$~Boo~A 
show a coherent period over the course of four observing seasons. 
Certainly, this activity is unusually stable.

\subsection{Determination of $v \sin i$ and $i$}

We determined the projected stellar rotational velocity ($v \sin i$) for each 
of these
stars using the  radial velocity ``template'' spectra (observed without the 
I$_{2}$ cell in place) obtained during the Keck observing runs. Using 
the newest 
version of the LTE line analysis code MOOG \citep{Sn73}, we first derived 
stellar 
parameters\footnote{A detailed analysis of the determination of stellar 
parameters and
abundances is provided in \citet{PaSnCo03}. A shortened description
is provided here.}- effective temperature ($T_{\rm eff}$), surface gravity 
(log $g$), 
microturbulence ($\xi$) and metallicity ([Fe/H]) for these four stars 
with interpolated\footnote{Interpolation software 
was kindly supplied by McWilliam (1995, private communication) and updated by 
Ivans (2002, private communication).} atmosphere models
based on the 1995 version of ATLAS9 code \citep{CaGrKu97}. We used no convective
overshoot in the model atmospheres.
We measured equivalent widths of about 20 unblended 
Fe I lines and 10 unblended Fe II lines for each star in the region 4490 to 
6175\AA (linelist provided in Paulson et al. 2003. In a self-consistent manner, $gf$ values for each line were derived 
from the Kurucz solar atlas \citep{KuFuBr84} and confirmed with a solar 
spectrum taken through HIRES. 
The stellar parameters 
derived are listed in Table 4. The average [Fe/H] values listed are relative to solar 
(log $\epsilon$(Fe/H)$_{\odot}$ = 7.52, \citet{SnKrPr91}). Using these 
parameters, we then synthesized 
a spectral region with 5 Fe I lines in the region 6150 - 6180\AA. In order
to determine $v \sin i$, we compute a disk intensity profile and then convolve
that with a broadening function. This broadening function contains not only 
$v \sin i$, but also a macroturbulent velocity and an instrumental broadening. 
We also incorporate into it a limb darkening coefficient.
We used a Gaussian profile to fit the lines of the thorium-argon (ThAr) lamp. 
We found a FWHM of 0.09\AA\ defines the instrumental 
broadening for this spectral region. The FWHM 
varies from 0.0918 to 0.0921\AA\ from the redmost to the 
bluemost lines, and the synthesis code is insensitive to this small a change. 
We estimated macroturbulence ($\zeta$) according to \citet{SaOs97} for
active stars and using B-V from \citet{APLa99}. Using estimates of limb 
darkening from \citet{Gr92},
were able to derive $v \sin i$ to within about 0.7~km~s$^{-1}$.

We estimated stellar radii ($R_{\star}$) from \citet{Gr92} using 
derived $T_{\rm eff}$, 
although we acknowledge that the stellar radii will be somewhat 
increased by
the higher metallicity of the Hyades.
Metal enrichment 
increases the opacity in the convection zone. According to hydrostatic 
equilibrium, as opacity is increased, the change in pressure as a 
function of optical depth will decrease, causing a slightly larger 
radius.  For the purposes here, an estimate of $R_{\star}$ based on solar 
metallicity will suffice.

Using $P_{\rm rot}$ measured from 
this work, we were able to estimate the rotational axis inclination ($i$) of 
each star. The sin$i$ values  are listed in Table 4. 
Adopting generous errors of 0.3 days for $P_{\rm rot}$ and 0.05 R$_{\odot}$ 
for $R_{\star}$, and model dependent errors of 100 K for $T_{\rm eff}$, 
0.3 km s$^{-1}$ for $\zeta$ and 0.1
km s$^{-1}$ for $\xi$, we note that the determination of sin$i$ is 
known to within 17\% for HD~26736 and HD~26767 and 24\% for HIP 13806.

The value of $i$ is useful in indicating the most
probable plane for planetary orbits, flagging
possible planet occultation candidates (though errors on $i$ make this a
blunt tool), and giving the 
general orientation of the star (guiding first-guess positions
for the location of active regions). 

\section{Modeling $v_{\rm r}$ variation due to spots and plage}

\subsection{Simple estimates of $v_{\rm r}$ variation due to spots}

It is interesting first to compare our results with prediction from the models 
of \citet{SaDo97}. The $v_{\rm r}$ amplitude ($A_{\rm s}$) due to a 
single spot goes as $A_{\rm s}\approx 6.5f_{\rm s}^{0.9}v{\rm sin}i$, 
where $f_{\rm s}\approx 0.4\Delta{\rm V}$ if we assume spot latitude of 
0$^{\circ}$, $\sin i =1$, 
and an average limb darkening coefficient of 0.6. $\Delta$V is 
the photometric amplitude in the V filter (here, we use the photometric 
amplitude in the Str\"omgren $y$ filter as an approximate substitute). 
We list $f_{\rm s}$ and $A_{\rm s}$ in Table 4. 
Figure 3 shows our results for $A_{\rm s}$,
the predicted amplitude of $v_{\rm r}$ due to spottedness, versus 
the observed $K$ velocity amplitude. Note that the $A_{\rm s}$ 
values are about a factor of two lower than $K$  
($\langle K/A_{\rm s}\rangle = 2.0 \pm 0.3$).
A very small part of this disagreement may come from the use of 
the $y$ filter as opposed to the V filter, but
we believe most of difference arises for two reasons. First, the \citet{SaDo97} 
spot models we used predict the maximum contribution
$v_{\rm r}$ for a given $f_{\rm s}$ 
(due to a single, equatorial spot on a $\sin i =1$ star). 
Considering multiple spots with the same total $f_{\rm s}$ or
a different $\sin i$ (\S 3.2) will only reduce the predicted 
$A_{\rm s}$ further.  However, the $f_{\rm s}$ values we used
were {\it also} derived assuming single spots; thus the true
$f_{\rm s}$ may be larger than Table 4 suggests. In the general case of multiple
spots, one must model the full $\Delta y$ and $v_{\rm r}$ curves.
Beyond this, the presence of significant amounts of plage on our 
targets (as implied by, e.g., Ca {\sc ii} HK emission; \citep{RaThLo87}), 
indicates another possible source of $v_{\rm r}$ fluctuations.
Thus, we must consider the possibility of multiple spots and plage
contributing to the complex $v_{\rm r}$ curves.

Based on the above argument, in the case of multiple spots, relations
connecting the rms {\it scatter} in $\Delta y$ and $v_{\rm r}$ might be more
useful than the amplitude.
Although the data are still limited, there appears to
be a trend between the rms scatter in 
$\Delta y$ ($\sigma_{y}$) and the rms $v_{\rm r}$ ($\sigma_{v}$). Figure 4 
shows $\sigma_{v}$ versus $\sigma_{y}$ of the observed data, including 
HD~166435, HD~19632 and HD~192263. 
A simple linear least squares fit yields 
$\sigma_{v}$[m s$^{-1}$] =3600$\times \sigma_{y}$+2.29. 
Since when $\sigma_{y}$=0 this implies $\sigma_{v} \approx 2.3$ m s$^{-1}
\approx \sigma_i$, the internal error of the $v_{\rm r}$ data, the fit
further supports the idea that much of the $\sigma_{v}$ in these stars is
  due to spots.

\subsection{Modeling the $v_{\rm r}$ effects of spots and plage}

The effects of dark starspots on the measured $v_{\rm r}$ have been
studied by \citet{SaDo97}, \citet{Ha99}, and \citet{Ha02}.  Plage,
areas of relatively strong magnetic fields and activity which are
optically bright, pose greater difficulties for modeling.  Unlike
spots, where the dominant effect is simply a strong reduction in the
local continuum, in plage the alteration of normal convective motions
by strong magnetic fields is not hidden from view by low surface
brightness.  To model the effects of plage we use the models presented
in \citet{Sa03}.  Briefly, we use observed solar bisectors taken in
plage and quiet regions at several limb angles as proxies for the
bisectors of stellar intensity profiles, $I_\nu$.  These proxies were
then used to ``warp" and shift symmetric $I_\nu$ profiles computed in a
simple Milne-Eddington model atmosphere.  We then employed the now
asymmetric quiet and plage $I_\nu$'s to construct model stellar flux
profiles for stars with any desired $v \sin i$, orientation, and plage
geometry.  Spots were modeled similarly, except that profiles inside
spots were assumed to be symmetric. The centroid of the resulting
profile was used to define $v_{\rm r}$ for the model.

We do not intend to present here a rigorous analysis of the spot and plage
contributions to $v_{\rm r}$; this is left for a future paper.  Our aim
is only to (1) show that simple spot/plage models can explain details of the 
RV, $y$ and H$\alpha$ variation beyond merely their amplitude and rms; and 
(2) argue that it is thus {\it plausible} that spot and plage
contribute significantly to the observed $\Delta v_{\rm r}$ variations.
Our best fit solutions are not unique, and only indicative of a class
of viable solutions under our chosen set of (hopefully physically reasonable)
simplifying assumptions.  
Since the data are also taken over an extended period ($\sim$4 months), the
best fit parameters necessarily represent time-averages of the related physical 
properties.  This is not entirely unrealistic: active longitudes on active 
stars are
known to persist for years (e.g., Jetsu 1996) and even solar active longitudes 
can be 
quite long-lived (Berdyugina \& Usoskin 2003).  
Time averages therefore do have some relevant physical meaning.
Since the plage models are based on solar line bisectors, we focused on
the most solar-like of our Hyad targets, HD~26736.  

Our assumptions are as follows. The most significant
one is that the spot/plage reside at latitudes $\theta =
30^\circ$ (near the subobserver $\theta \approx 20^\circ$ for HD~26736),
which yields close to the maximum effect for a given region area.  We
adopt a spot temperature of $T_s = 4000$ K (similar to the Sun), corresponding 
to a
fractional continuum difference in spots at 5000\AA\ (relative to the quiet
photosphere) of $\Delta I_{S}= (I_{Q}-I_{S})/I_{Q}$ = $-$90\%. Additionally, 
for simplicity, we took the fractional continuum difference in plage
(also relative to the quiet photosphere), to be
$\Delta I_{P}=(I_{Q}-I_{P})/I_{Q}=0$.  A linear limb-darkening
coefficient $\epsilon = 0.6$ was adopted, and the regions were assumed
circular.

We first attempted to model the observed $\Delta y$ from HD~26736, successively
adding features of varying radius and longitude to fit $\Delta y$.  Since we
have adopted $\Delta I_P = 0$, spots must completely account for the
observed $\Delta y$.  We then apply the resulting spot sizes and positions to
compute a model for spot-induced $v_{\rm r}$ variation.  By not {\it
fitting} $v_{\rm r}$, but rather letting the observed $\Delta y$ drive the
modeling, we ensure that the resulting $v_{\rm r}$ model is consistent
with $\Delta y$ but does not attempt to ``explain" features which are not due
to spots.  For our assumed $\theta$, a single spot spans only a phase
range $\delta \phi \approx 0.5$ (Fig. 5a) and thus cannot alone
describe the photometry.  The $\Delta y$ variation could be reasonably well
described with two features (see Table 5 for their properties; fit
RMS $\sigma_{\rm fit}(y)$ = 0.00394)
We find an optimum ``background" unmodulated brightness of $\Delta y = 0.988$
(due to the pristine photosphere plus any uniform spot component).
The resulting $v_{\rm r}$ model agrees fairly well for $\phi > 0.85$ and
$\phi < 0.4$, but there are some discrepancies elsewhere.
The discrepancies are strongest around phase $\phi = 0.75$, where the
differences  between $v_{\rm r}$ and the spot-only model are $\approx
50$ m s$^{-1}$.  Adding a third spot could partly correct this, only to
introduce new fitting errors to $v_{\rm r}$ around $\phi = 0.9$, and
errors in $\Delta y$ near $\phi = 0.6$.

The remaining systematic differences between the spot-only model and
$v_{\rm r}$ could be due to plage, our assumptions (especially the
restriction on $\theta$), or even potentially a planet.  To investigate
the first possibility, it is useful to have a plage diagnostic
analogous to $\Delta y$.  As our HET spectra do not contain the traditional 
plage
indicator Ca {\sc ii} H \& K, we constructed a substitute from the
H$\alpha$ profile as follows.  First, telluric features were removed by
ratioing the HD~26736 data with a scaled spectrum of a rapidly rotating A
star, whose broad H$\alpha$ feature was removed with a cubic spline
fit.  The average flux in a 2.15\AA~interval centered on the H$\alpha$
core was then ratioed with the average flux in a nearby, nearly
line-free ``continuum" segment (5\AA~centered at 6602\AA) to form an
H$\alpha$ index $S_{{\rm H}\alpha} = F_{{\rm H}\alpha {\rm
core}}/F_{\rm con}$. This index showed a small but significant
modulation at $P_v$ (Fig.~5b).  Comparison of Figs. 5a and 5b
reveals that the plage emission areas are not coincident with the
spots: the $S_{{\rm H}\alpha}$ curve maximum is shifted from the $\Delta y$
minimum, and there are features in $S_{{\rm H}\alpha}$ not obvious in
$\Delta y$ (e.g., the enhancement near $\phi \approx 0.9$).

Next, in a fashion similar to the spot-modeling we added plage 
regions until the  the $S_{{\rm H}\alpha}$ data was reasonably well 
fit. We assumed that the H$\alpha$ emission is limb-brightened 
with $\epsilon = -0.2$. An estimate for the intrinsic plage emission strength
per unit area, $I_{S({\rm H}\alpha)}$, was also needed.  There 
has been relatively little work using H$\alpha$ as an activity diagnostic 
in low-to-moderate activity G and K stars. \citet{He85}
found the minimum flux for an early G star 
in a 1.7\AA~bandpass under the H$\alpha$ core
(expressed as an equivalent width) is $W_\lambda \approx 0.74$ \AA. 
His most active target, $\xi$ UMa B (G5V, $P_{\rm rot}= 3.98$d) observed with
a slightly different bandpass, showed $\Delta W_\lambda \approx 0.277$ 
\AA~above this baseline level.
If we assume the latter represents a ``saturated" chromosphere
star ($f_p \approx$1) and the former a completely inactive star ($f_p \approx
0$), we find (after correcting for resolution and bandpass differences) 
$I_{S({\rm H}\alpha)} \approx 0.016$ for a plage at disk center with an
area of 10\% of the visible surface.  With these assumptions we found
good fits came for a ``background" $S_{{\rm H}\alpha} = 0.475$ (due to
the photospheric H$\alpha$ and any uniform plage/network component).
We note, however, that there is a trade-off between plage brightness
and area, i.e., fits to $S_{{\rm H}\alpha}$ are equivalent for
$I_{S({\rm H}\alpha)} \times \Sigma f_p \approx$ constant.  Thus,
changes in $I_{S({\rm H}\alpha)}$ affects plage areas and thereby the
plage contribution to $v_{\rm r}$.
A minimum of 3 regions were required (see Table 5; $\sigma_{\rm
fit}(S_{{\rm H}\alpha})$ = 0.00219); the positions of the model spots and plage
are also indicated in Fig. 5.  The resulting plage-induced $\Delta v_{\rm r}$
ends up being rather small relative to the spot contribution (model
amplitudes of $A_P \approx 16$ m s$^{-1}$ compared with $A_S \approx
83$ m s$^{-1}$; Fig. 5d).  The inclusion of the plages (at $\theta =
30^\circ$) does not significantly alter the agreement between the
resulting $v_{\rm r}$ model and the data: the RMS between them is
$\sigma$ = 29.1 m s$^{-1}$.  Including plage makes the agreement
slightly worse around $\phi \approx 0.1 - 0.3$ by the plage, but
slightly better near $\phi \approx 0.5 - 0.7$; significant
discrepancies remain (Fig. 5d).  Whether these discrepancies show any
systematic trends (suggesting a possible underlying planetary signal)
must await more rigorous modeling, in particular relaxing the
assumption of $\theta = 30^\circ$ for all regions. We leave this for a
future paper.

\section{Discussion and Conclusions}

This initial search for short-period planets in the Keck Hyades survey
has instead turned up several lines of evidence pointing to $v_{\rm r}$
variations driven by magnetic activity.  First, the
photometric and the $v_{\rm r}$ variations show similar periodicities (Fig. 1; 
Table 1);
the one star studied here in H$\alpha$ (HD~26736) shows a similar periodicity in
that activity diagnostic as well (Fig.~5b).  The $v_{\rm r}$
amplitudes increase with the $\Delta y$ photometric amplitudes in a way
consistent with (though smaller than) predictions of a simple
single-spot model \citep[][ Fig. 3]{SaDo97}.  The rms scatter in $\Delta y$ and
$v_{\rm r}$ also show a linear relationship (Fig. 4). This correlation
is  potentially quite useful as a simple way of estimating $\sigma_v$
from photometry.  More data must be collected before we can predict
$\sigma_v$ from $\sigma_y$ with confidence; the relationship may depend
on other properties/parameters.  Once refined, though, such a
correlation may be of use to rapidly help flag ``problem" stars which
will require more careful analysis to confirm planets, to screen out
such stars from search lists, or to estimate what fraction of
a given star's $\sigma_v$ is likely due to activity.  

The phase shifts between the $v_{\rm r}$ and $\Delta y$ curves clearly seen in
at least two of the stars (HIP~13806 and HD~26736) are also consistent with a
spot origin for much of the $v_{\rm r}$ variation.  To see this,
consider that the perturbation from a single, black spot at $\theta=0$
on a star (with $i = 90^\circ$ and no limb-darkening) scales as
$\Delta v_{\rm r} \propto - f_s \sin \phi \cos \phi$ (where $\phi$ is
the phase angle measured from the sub-observer meridian $\phi=0$) while
photometry varies as $\Delta y \propto f_s \cos \phi$. The minimum
light (at $\phi =0$) is shifted from the maximum $\Delta v_{\rm r}$ (at
$\phi = -45^\circ$); the actual shift will depend on details of the
limb-darkening and active region geometry.  HIP~13806 and HD~26736 (Fig. 1)
both show this pattern of $v_{\rm r}$ maxima preceding light minima.
The case of HD~26767 is less clear (two $v_{\rm r}$ near $\phi \approx
0.4$ are discrepant), but if one identifies the primary maximum with
the feature near $\phi \approx 0.2$, it also follows the pattern.
\citet{QuHeSi01} also see this phase shift in HD~166435.

Finally, beyond these more qualitative connections of $v_{\rm r}$ with
spots and plage, we have also demonstrated that the $v_{\rm r}$ variations
can be modeled directly.  By first modeling the observed photometry and
H$\alpha$ emission and using the inferred spot and plage locations and
sizes in simple models of these regions' $v_{\rm r}$ properties,  we
can explain a significant fraction of the $v_{\rm r}$ variation of vB
15 in detail (Fig. 5).  Given internal errors of $\sigma_i(y) \approx
0.003$ (the constancy of the comparison stars) and $\sigma_i(S_{{\rm
H}\alpha}) \approx 0.0025$, our fits to $\Delta y$ and $S_{{\rm H}\alpha}$ are
reasonably successful with two spots and three plages, respectively
($\chi^2_\nu = 1.7$ and $\chi^2_\nu = 0.8$).  The resulting $v_{\rm r}$
models are less successful in explaining all the velocity variations
($\sigma \approx 29$ m s$^{-1}$; $\chi^2_\nu \approx 26$)
but is able to account for $\approx$50\% of the variance in the $v_{\rm r}$
data.  Some of the discrepancies are undoubtedly due to our
simplifying assumptions, such as all regions placed at $\theta =
30^\circ$. Some less obvious implicit assumptions (e.g., that all spots
and all plages are identical in how they act on $v_{\rm r}$, that the
solar plage bisectors used as proxies are representative of stellar
plage) may also be important.  Still, despite the fact that the
modeling presented here is simplified and certainly not definitive, we
believe that it argues that a combination of spot and plage can explain
many of the $v_{\rm r}$ fluctuations seen in HD~26736. By analogy, we suspect 
that many of the $v_{\rm r}$  variation seen in the other targets can be 
similarly explained by activity. Indeed, the $v_{\rm r}$ ``jitter'' in HD~26756
without corresponding $\Delta y$ changes or strong periodicity might be 
the result of rapidly evolving plage dominating the $v_{\rm r}$ perturbations.

Young stars present distinct problems for the search for short-period
planets. Our work shows that photometric confirmation along with good
activity measurements are a very helpful check to insure the viability
of short-period planetary candidates.  It is important to note that
active stars need not necessarily be excluded from planet searches
since it should be possible (at least in principle) to remove
activity-related perturbations from $v_{\rm r}$ measurements.  \citet{SaFi00}
show that simple correlations between $v_{\rm r}$ and
a plage diagnostic (e.g., H$\alpha$, Ca {\sc ii} H \& K) can be
effective in removing long-timescale variations in $v_{\rm r}$
stemming, for example, from magnetic cycle variations in mean plage
area.  On shorter (rotational) timescales, this might be accomplished
by modeling the activity features (as explored here) or by monitoring
line shape changes (e.g., bisectors) as indicators of rotating
inhomogeneous features \citep{SaFiSn01, SaHaCo02, QuHeSi01}.  

Magnetic activity and hence activity-induced $v_{\rm r}$ variations will 
occur on a wide range of 
timescales (see for example, Fig. 11 from \citet{Je02} showing solar
photometric variability extrapolated to stars; also, \citet{DoDoBa97a,DoDoBa97b}
for stellar Ca {\sc ii} HK variability). It is important to note,
however, that 
most of the strongly {\it periodic} power will be concentrated on 
distinct surface/activity timescales: $P_{\rm rot}$, differential rotation,
active region growth/decay, active longitude growth/decay (including 
``flip-flops"; see e.g., \nocite{JePeTu93}Jetsu, Pelt, \& Tuominen 1993), and 
magnetic cycle.  Planets with orbital periods well 
separated from these timescales will be much easier to detect 
and confirm.

The models also indicate the possibility that plage-induced $v_{\rm r}$
fluctuations without strong, parallel photometric variations 
must also be considered as possible contributors
to $v_{\rm r}$ signals.   Since plage-to-spot area ratios are
largest on inactive stars, it is quite possible that plage-induced
$\Delta v_{\rm r}$ may dominate on these objects.

Quite apart from planet searches, our work 
suggests the possible use of precision $v_{\rm r}$
measurements to investigate surface 
features on cool stars. The $v_{\rm r}$ variation
as a function of rotational phase is distinctly different for
spots and plage, and shows significantly sharper changes with $\phi$ than
photometry or chromospheric activity (Fig. 5c).  This makes
high precision $v_{\rm r}$ curves a powerful tool for
investigating stellar surface structures, and one uniquely suited
for the study of plages and slower rotators ($v \sin i \la 12$ km s$^{-1}$), 
for which Doppler imaging is less useful.

\acknowledgments
DBP and WDC are supported by NASA grant  
NAG5-9227 and NSF grant AST-9808980. 
SHS was supported by NASA Origins program grant NAG5-10630.
GWH acknowledges support from NASA 
grants NCC5-96 and NCC5-511 as well as NSF grant HRD-9706268. 
We would like to thank Artie Hatzes  and Chris Sneden
for many useful discussions and Rob Robinson and Frank Bash for 
making the time critical observations at HET possible.


\newpage
\begin{figure}
\plotone{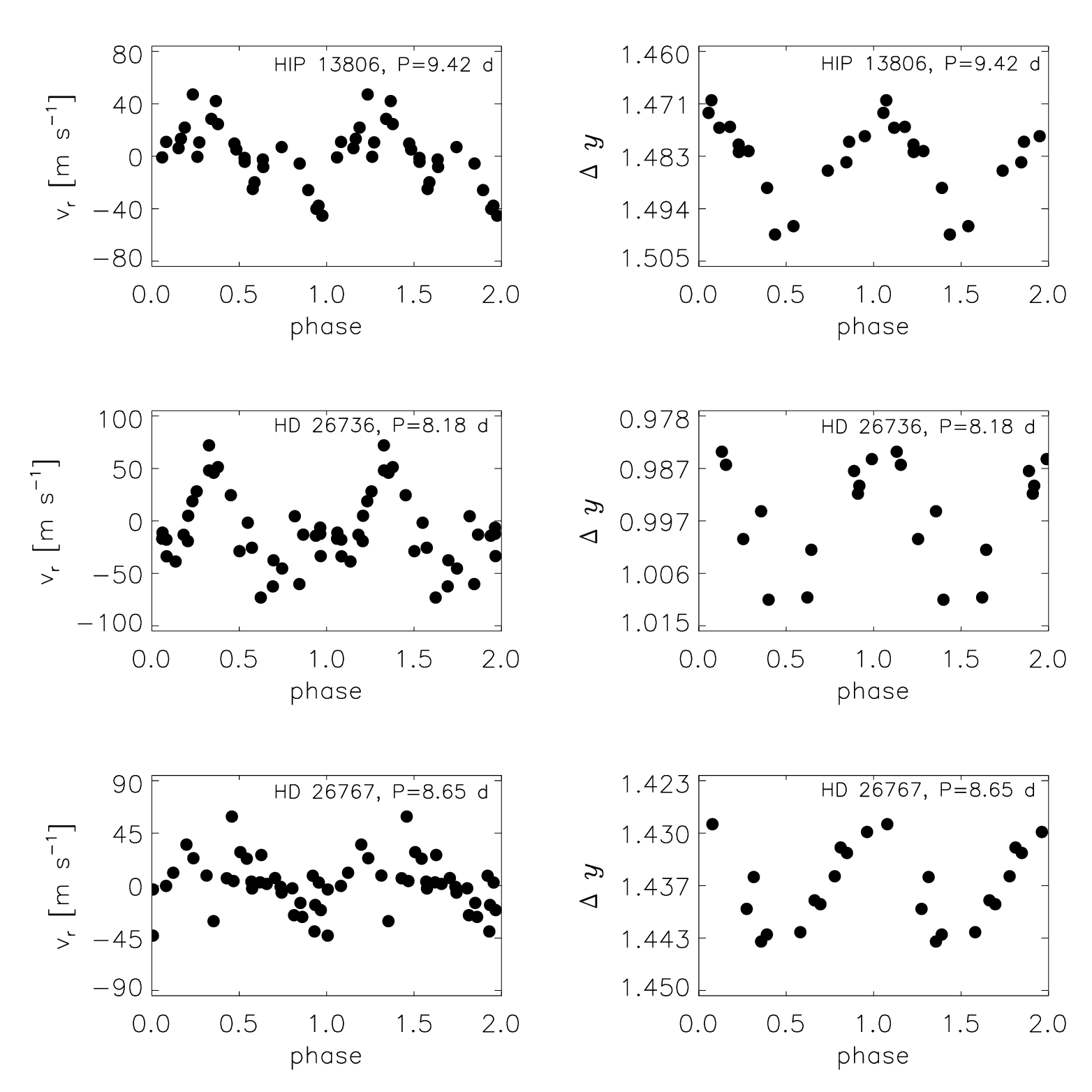}
\caption{Program stars: $v_{\rm r}$ (left) and differential Str\"omgren $y$ 
(right) curves.} 
\end{figure}

\begin{figure}
\plotone{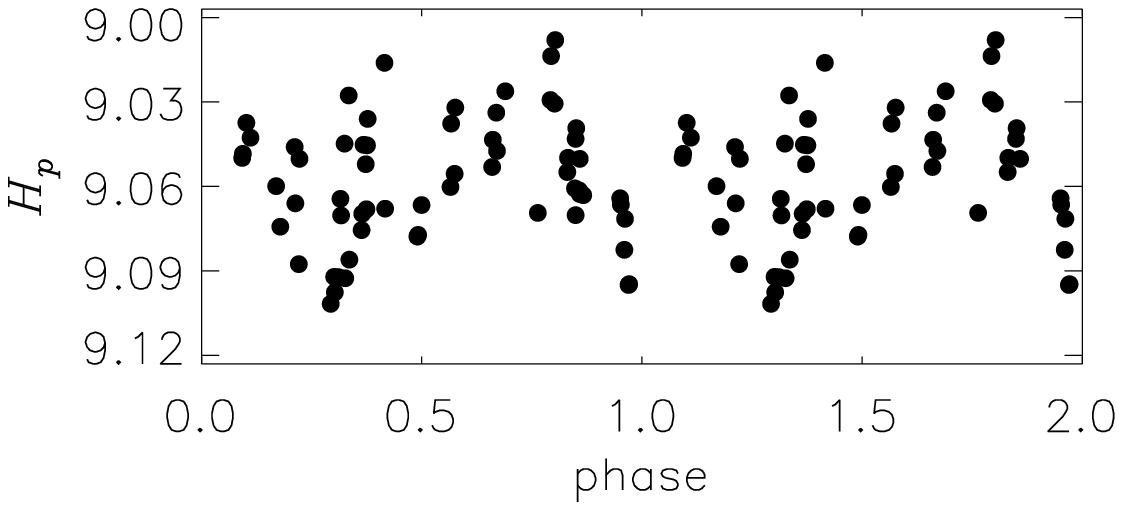}
\caption{Hipparcos photometric data ($H_{\rm p}$) for HIP~13806, phased to 9.42 
days.}
\end{figure}

\begin{figure}
\plotone{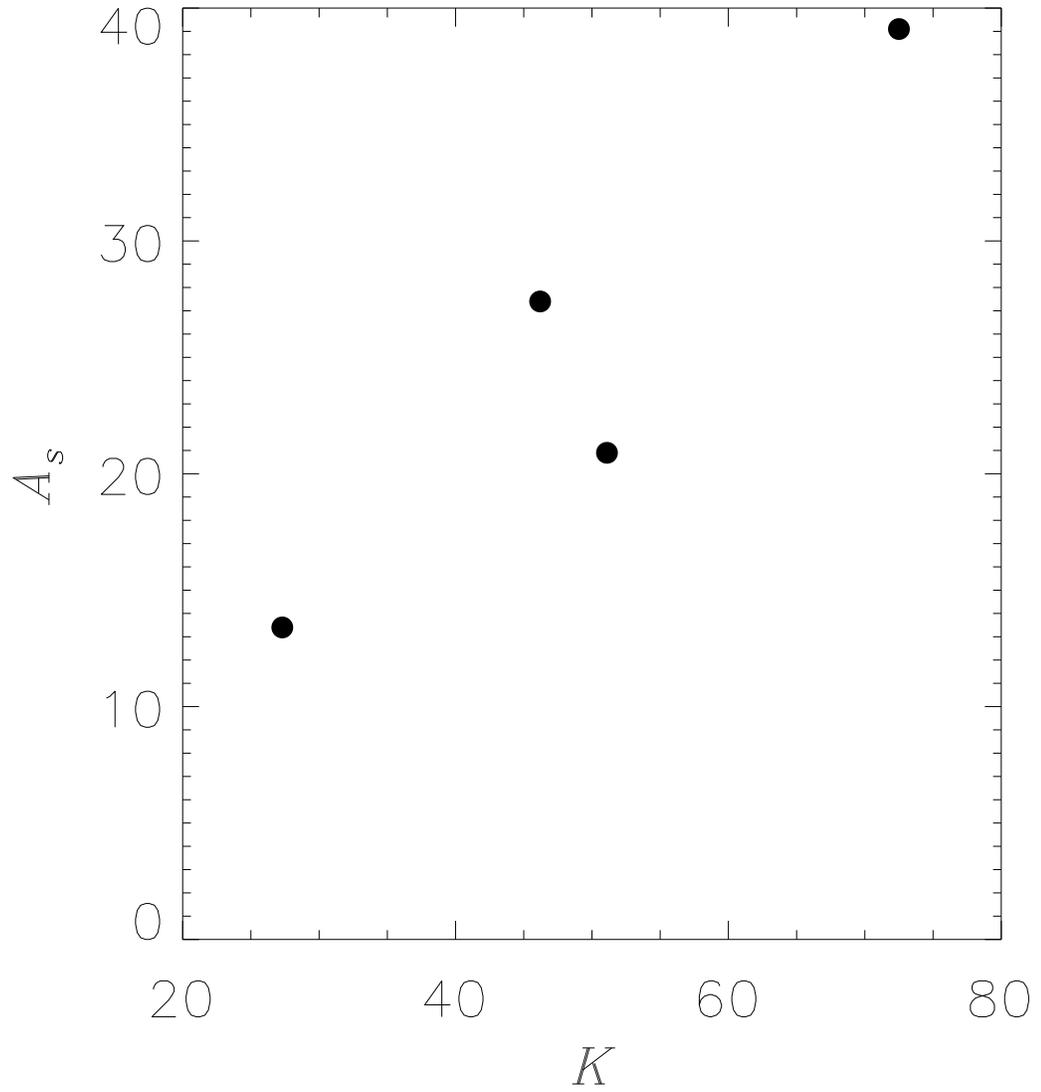}
\caption{$v_{\rm r}$ amplitude, $A_{\rm s}$, as predicted from \citet{SaDo97}
versus $K$ for our program stars.}
\end{figure}

\begin{figure}
\plotone{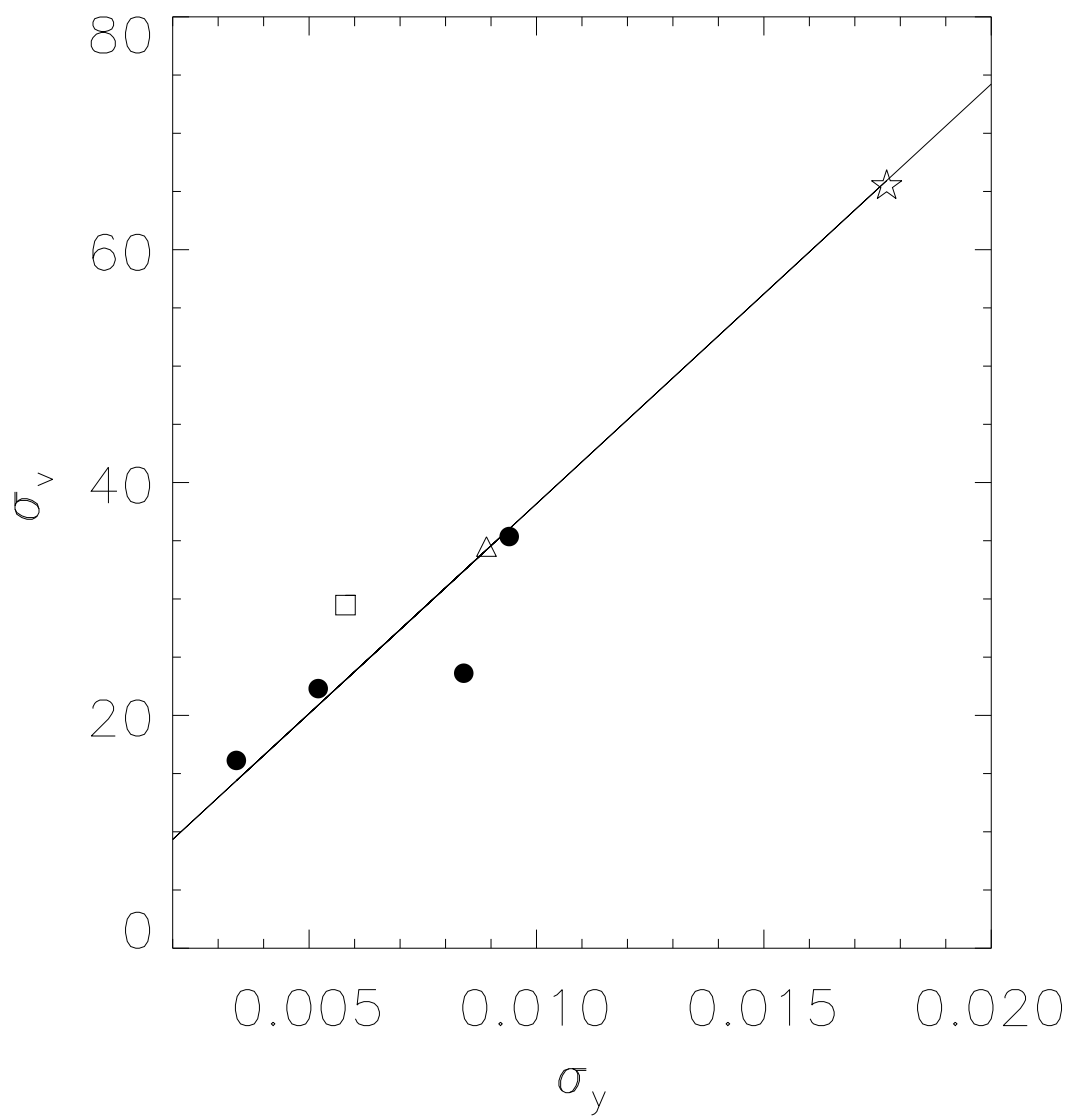}
\caption{$v_{\rm r}$ rms versus $y$ rms for our program stars ($circles$),
HD~166435 \citep{QuHeSi01} ($star$), HD~19632 (private communication, G. Henry, 
P. Butler 2002) ($box$) and HD~192263 \citep{HeDoBa02} ($triangle$).}
\end{figure}

\begin{figure}
\plotone{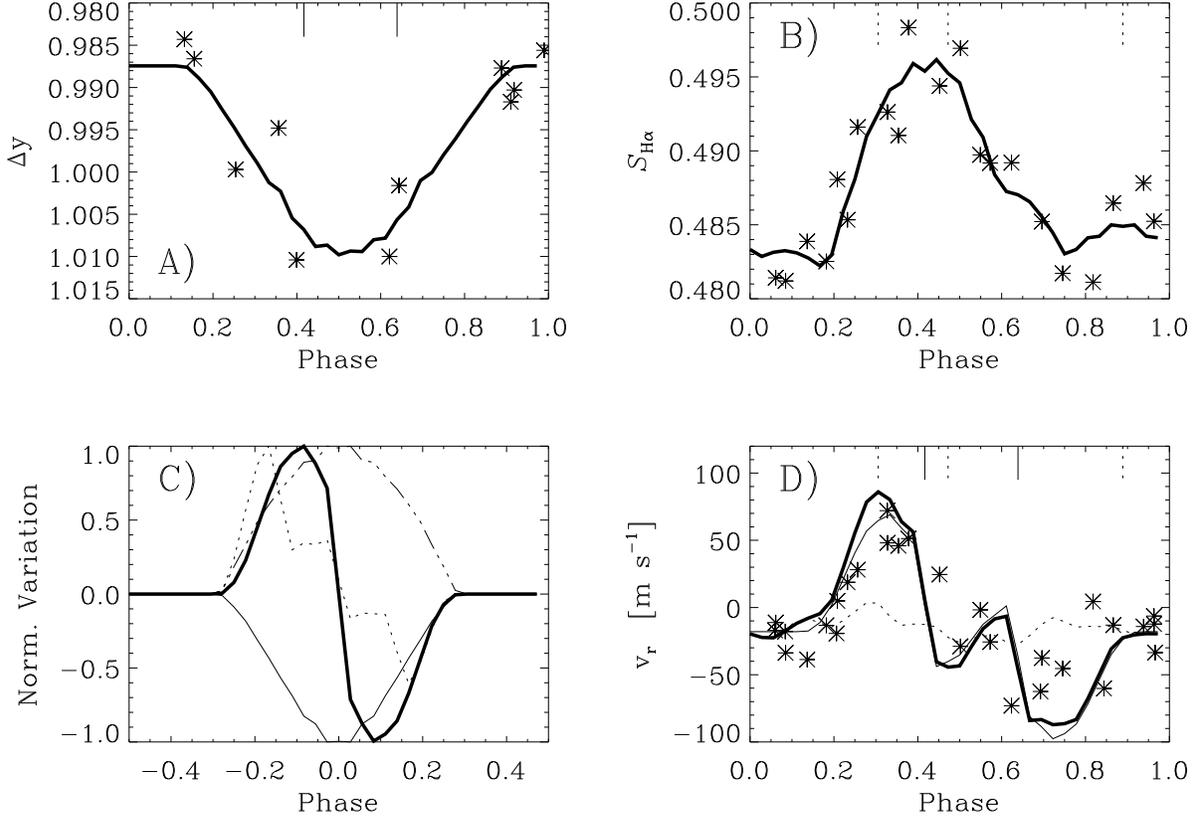}
\caption{Observed $\Delta y$, $S_{{\rm H}\alpha}$, and $v_{\rm r}$ for HD 26736 
plus models (see Table 5 for model parameters, text for details). 
A) Observed $\Delta y$ (stars), best-fit two spot model for $\Delta y$ 
(solid; $\sigma_{\rm fit} = 0.00394$), with 
phase of model spot meridian passage marked (vertical ticks).
B) $S_{{\rm H}\alpha}$ computed from observed $H_\alpha$ profiles 
(stars), best-fit 3 plage model for $S_{{\rm H}\alpha}$ (solid;
$\sigma_{\rm fit} = 0.00219$), with 
phase of plage meridian passage marked (vertical ticks).
C) Normalized variation for a single spot $v_{\rm r}$ (heavy solid), 
$\Delta y$ (solid), single plage $v_{\rm r}$ (heavy dashed), and 
$S_{{\rm H}\alpha}$
(dashed). Maximum amplitudes are 25.5 m s$^{-1}$ and 0.0112 mag 
for a $f_S$=1\% spot, and 3.0 m s$^{-1}$ and 0.0128 H$\alpha$ units 
for a $f_P$=1\% plage.   
D) Observed $v_{\rm r}$ (stars), $v_{\rm r}$ of the 
two spot model which best fits $\Delta y$ in panel A (solid; 
$\sigma = 29.0$~m~s$^{-1}$), $v_{\rm r}$ of the 3 plage model which 
best fits
$S_{{\rm H}\alpha}$ in panel B (dotted), and the $v_{\rm r}$ of the 
combined spot + plage model (heavy solid; $\sigma = 29.1$ m s$^{-1}$). 
Vertical ticks (top) mark phase of central meridian passage of model spots 
(solid) and plages (dashed).  
}
\end{figure}

\newpage

\clearpage
\begin{deluxetable}{lcccccccccccc}
\tabletypesize{\scriptsize}
\tablecaption{Program Information \label{tbl-1}}
\tablewidth{0pt}
\tablehead{
\colhead{Star} & 
\colhead{vB\#} & 
\colhead{$\# v_{\rm r}$}   &
\colhead{K} & 
\colhead{P$_{rot,v_{r}}$} & \colhead{FAP$_{v_{\rm r}}$} &
\colhead{$\#$ phot.} &
\colhead{$\Delta y$} & 
\colhead{P$_{rot,phot}$} & \colhead{FAP$_{y}$}&
\colhead{$\#$ Hipp.} &
\colhead{$P_{\rm rot, Hipp.}$}& \colhead{FAP$_{\rm HIP}$}\\
\colhead{}& &obs. &[m s$^{-1}$]&[days]& [\%] &
obs. & &[days] &[\%] & obs.& [days] & [\%] }
\startdata
HIP 13806 & 153 & 25 & 46.2 & 9.42& 1.7 & 14 & 0.028 & 9.18 & 4.3 &62&9.60& 
0.02\\ 
HD 26767& 18   & 30 & 51.1 & 8.65 & 7.9 &12 & 0.015 & 8.65& 5.9& 34&
\nodata&\nodata\\
HD 26736& 15   & 29 & 72.5 & 8.18 & 1.4 &11 & 0.026 & 8.44& 15.5&102&
\nodata&\nodata\\
HD 26756& 17   & 24 & 27.3 &\nodata & \nodata & 12 & 0.011&\nodata& 
\nodata &44&\nodata&\nodata\\
\enddata
\end{deluxetable}

\clearpage
\begin{deluxetable}{lccc}
\tabletypesize{\scriptsize}
\tablecaption{Radial Velocities \label{tbl-2}}
\tablewidth{0pt}
\tablehead{
\colhead{Star} &
\colhead{JD - 2400000}&
\colhead{$v_{\rm r}$ [m s$^{-1}$]} & \colhead{Uncertainty [m s$^{-1}$]}}
\startdata
HIP 13806&52262.770111  &    27.11 & 6.35       \\
&52263.768303  &     8.46 & 5.56       \\
&52265.558250  &   -21.27 & 6.61       \\
&52266.564679  &   -28.23 & 6.80       \\
&52269.747800  &   -65.33 &  18.06     \\
&52270.756446  &    -9.11 & 6.52       \\
&52271.754419  &     1.83 & 8.31       \\
&52297.687220  &   -60.20 & 4.61       \\
&52299.682102  &   -13.88 & 4.89  \\
&52300.694922  &   -20.44 & 4.46  \\
&52301.692142  &    22.06 & 7.63  \\
&52302.690889  &   -10.32 & 3.82  \\
&52303.678815  &   -44.98 & 4.75  \\
&52306.669150  &   -45.84 & 3.63  \\
&52312.664745  &   -24.29 & 7.08   \\
&52313.641222  &   -22.49 & 3.22   \\
&52314.656157  &   -13.03 &  16.02 \\
&52315.627230  &   -25.59 & 5.16 \\
&52316.642697  &   -57.75 & 4.17 \\
&52317.638826  &   -20.89 & 5.27 \\
&52318.648681  &    -6.70 & 5.12 \\
&52319.639116  &    -9.43 & 5.11 \\
&52320.642096  &     4.52 & 5.29 \\
&52321.632011  &   -14.87 & 5.60 \\
&52322.615422  &   -39.87 & 3.69 \\
HD 26767&52199.811224 &     59.23   &      5.84      \\
&52200.812501 &     -2.48   &      5.58     \\
&52202.803337 &     -2.36   &      4.92    \\
&52203.814494 &      8.44   &      6.58   \\
&52236.872727 &     -6.04   &      7.40  \\
&52237.893749 &    -26.90   &      6.69 \\
&52252.676137 &      3.40   &      9.22   \\
&52253.845854 &      6.45   &      5.58  \\
&52255.835060 &    -16.65   &      5.01\\
&52261.818689 &     26.24   &      7.64 \\
&52299.701994 &    -42.94   &      5.98  \\
&52300.710344 &     11.01   &      4.25   \\
&52301.708195 &     23.47   &      4.49    \\
&52302.706162 &    -30.54   &      4.56     \\
&52303.693759 &      3.89   &      3.78      \\
&52306.688347 &    -25.46   &      4.00       \\
&52307.697467 &    -39.37   &      4.80\\
&52312.681499 &     28.70   &      4.82 \\
&52313.660480 &      2.78   &      4.51  \\
&52314.671760 &     -1.10   &      4.39   \\
&52315.651757 &    -14.89   &      4.96    \\
&52316.661289 &    -21.04   &      4.07 \\
&52317.656991 &     -0.28   &      5.26  \\
&52318.663084 &     35.16   &      4.12   \\
&52319.657755 &      8.61   &      4.17    \\
&52320.659775 &      6.29   &      4.30     \\
&52321.649747 &     23.07   &      4.23\\
&52322.634036 &      1.43   &      4.13 \\
&52325.654987 &     -3.34   &      4.86  \\
HD 26736&52238.711713 &      2.63   &      5.79   \\
&52263.619768  &   -12.31 & 7.38       \\
&52264.606026  &   -17.85 & 5.86       \\
&52265.606621  &   -19.30 & 6.01       \\
&52266.597528  &    72.02 & 6.38       \\
&52269.584370  &   -62.45 & 7.75       \\
&52270.823349  &   -60.25 & 7.72       \\
&52271.821446  &   -33.54 & 7.08       \\
&52272.587749  &   -16.98 & 6.58       \\
&52297.754247  &   -38.71 & 6.04       \\
&52298.738493  &    28.13 & 4.79       \\
&52299.731856  &    51.27 & 5.45       \\
&52300.743482  &   -28.87 & 6.69       \\
&52301.739367  &   -73.07 & 9.49       \\
&52302.737520  &   -45.38 & 5.23       \\
&52303.724341  &   -13.08 & 4.96       \\
&52306.719880  &    18.78 & 5.27       \\
&52307.715564  &    45.90 & 4.02       \\
&52312.698359  &    -6.38 & 4.56       \\
&52313.693807  &   -33.77 & 3.38       \\
&52314.702460  &     4.90 & 4.38       \\
&52315.681787  &    48.11 & 5.03       \\
&52316.697637  &    24.51 & 4.73       \\
&52317.684034  &   -25.61 & 4.63       \\
&52318.693958  &   -37.56 & 5.22       \\
&52319.690237  &     4.46 & 4.76       \\
&52320.675927  &   -14.14 & 4.64       \\
&52321.680885  &   -11.19 & 6.62       \\
&52322.667651  &   -13.24 & 5.56       \\
&52325.671298  &    -1.74 & 4.41       \\
HD 26756&52200.796965  &   -11.69   &      6.23 \\
&52202.783532  &    25.82   &      6.74\\
&52203.989956  &    16.65   &      7.52   \\
&52247.879099  &    13.27   &      8.14  \\
&52249.861960  &   -19.19   &      6.70 \\
&52252.660236  &    -8.91   &      8.12\\
&52253.861596  &    -7.63   &      7.69   \\
&52255.851520  &    -5.29   &      6.40  \\
&52278.591680  &    16.13   &      8.26 \\
&52299.717028  &    10.40   &      5.72  \\
&52300.725215  &   -28.88   &      5.29     \\
&52301.721528  &   -18.92   &      5.71    \\
&52302.719726  &   -17.26   &      4.65   \\
&52303.709706  &     2.60   &      5.18      \\
&52306.703796  &   -26.66   &      4.96     \\
&52313.677015  &    15.59   &      4.35    \\
&52314.685147  &   -13.61   &      5.33   \\
&52315.666452  &     3.35   &      4.85  \\
&52316.679270  &    -1.12   &      5.27      \\
&52317.670296  &   -16.98   &      5.36     \\
&52318.677153  &     0.27   &      5.77    \\
&52319.673016  &    16.77   &      5.08   \\
&52321.664983  &    25.44   &      5.34  \\
&52322.651275  &    -1.53   &      5.26 \\
\enddata
\end{deluxetable}

\clearpage
\begin{deluxetable}{lccc}
\tabletypesize{\scriptsize}
\tablecaption{Photometry\label{tbl-3}}
\tablewidth{0pt}
\tablehead{
\colhead{Star} &
\colhead{JD - 2400000}&
\colhead{$\Delta y$}}
\startdata
HIP 13806&52314.6093 &1.4856\\
&52315.6109 &1.4838\\
&52316.6031 &1.4782\\
&52317.6049 &1.4732\\
&52327.6049 &1.4764\\
&52328.6443 &1.4800\\
&52328.6548 &1.4816\\
&52330.6054 &1.4993\\
&52331.5996 &1.4975\\
&52334.5989 &1.4794\\
&52336.5999 &1.4705\\
&52337.5993 &1.4762\\
&52338.5994 &1.4814\\
&52339.5992 &1.4893\\
HD 26767&52315.6295& 1.4323\\
&52316.6226 &1.4296\\
&52317.6234 &1.4286\\
&52328.6820 &1.4437\\
&52330.6221 &1.4425\\
&52331.6163 &1.4389\\
&52332.6166 &1.4316\\
&52336.6171 &1.4395\\
&52337.6166 &1.4428\\
&52345.6180 &1.4354\\
&52348.6278 &1.4384\\
&52349.6254 &1.4353\\
HD 26736&52328.6250 &0.9917\\
&52328.6897 &0.9903\\
&52330.6308 &0.9866\\
&52332.6253 &1.0104\\
&52334.6239 &1.0016\\
&52336.6245 &0.9877\\
&52338.6183 &0.9843\\
&52339.6184 &0.9997\\
&52342.6175 &1.0100\\
&52345.6350 &0.9856\\
&52348.6352 &0.9948\\
HD 26756&52315.6295 &1.8513\\
&52316.6226 &1.8556\\
&52317.6234 &1.8442\\
&52328.6820 &1.8514\\
&52330.6221 &1.8531\\
&52331.6163 &1.8516\\
&52332.6166 &1.8533\\
&52336.6171 &1.8450\\
&52337.6166 &1.8531\\
&52345.6180 &1.8487\\
&52348.6278 &1.8495\\
&52349.6254 &1.8502\\
\enddata
\end{deluxetable}

\clearpage
\begin{deluxetable}{ccccccccccc}
\tabletypesize{\scriptsize}
\tablecaption{Stellar Parameters \label{tbl-4}}
\tablewidth{0pt}
\tablehead{
\colhead{Star} &
\colhead{vB \#} &
\colhead{$T_{\rm eff}$} &
\colhead{log $g$} &
\colhead{$\xi$} &
\colhead{[Fe/H]}&
\colhead{$\zeta$} &
\colhead{$v \sin i$}&
\colhead{sin$i$} & \colhead{$f_{\rm s}$} & 
\colhead{$A_{\rm s}$}\\
\colhead{}&&[K]&&[km s$^{-1}$]&&[km s$^{-1}$]&[km s$^{-1}$]&&[\%]&[m s$^{-1}$]
}
\startdata
HIP 13806 &vB 153 & 5150 & 4.5 & 0.60 & 0.19 & 2.3 & 3.8 & 0.91&1.10&27\\
HD 26767 &vB 18  & 5900 & 4.4 & 0.80 & 0.23 & 3.9 & 5.4 & 0.83&0.60&21\\
HD 26736 &vB 15  & 5750 & 4.4 & 0.80 & 0.19 & 3.8 & 5.4 & 0.94&1.00&39\\
HD 26756 &vB 17  & 5650 & 4.4 & 0.80 & 0.17 & 3.5 & 4.5 & \nodata&0.44&13\\
\enddata
\end{deluxetable}

\begin{deluxetable}{ccc}
\tabletypesize{\scriptsize}
\tablecaption{Spot and Plage Parameters: HD 26736 \label{tbl-5}}
\tablewidth{0pt}
\tablehead{
\colhead{Region } &
\colhead{$\phi_0$} &
\colhead{Area} \\ 
\colhead{} &  & [\%] 
}
\startdata
spot 1 & 0.42 & 1.15 \\
spot 2 & 0.64 & 1.05 \\
plage 1 & 0.31 & 1.15 \\
plage 2 & 0.47 & 2.30 \\
plage 3 & 0.89 & 1.30 \\
\enddata
\end{deluxetable}

\end{document}